\documentclass[letterpaper, 10pt, conference]{ieeeconf}      % Use this line for a4
\usepackage{graphicx} 
\usepackage{url}
\IEEEoverridecommandlockouts                              % This command is only
\usepackage{geometry}
\usepackage[colorinlistoftodos]{todonotes}
\title{\LARGE \bf
BrailleBand: Blind Support Haptic Wearable Band for Communication using Braille Language
}

\author{Savindu H.P.$^{1}$, Iroshan K.A., Panangala C.D., Perera W.L.D.W.P., De Silva A.C. (MIEEE)% <-this % stops a space
\thanks{*This work was supported by the University of Moratuwa, Faculty of Engineering.}% <-this % stops a space
\thanks{$^{1}$Savindu H.P. is with the Department of Electronic and Telecommunication Engineering, Faculty of Engineering, University of Moratuwa, 10400 Katubedda, Sri Lanka.
        {\tt\small savinduherath@gmail.com}}%
}

\begin{document}

\maketitle
\thispagestyle{empty}
\pagestyle{empty}

%%%%%%%%%%%%%%%%%%%%%%%%%%%%%%%%%%%%%%%%%%%%%%%%%%%%%%%%%%%%%%%%%%%%%%%%%%%%%%%%
\begin{abstract}

%Assistive devices developed through assistive technologies have become the key applications of biomedical engineering which lead to the enhancement of the quality of life of people with physical and sensory disabilities and abnormalities. 
Visually impaired people are neglected from many modern communication and interaction procedures. Assistive technologies such as text-to-speech and braille displays are the most commonly used means of connecting such visually impaired people with mobile phones and other smart devices. Both these solutions face usability issues, thus this study focused on developing a user friendly wearable solution called the `BrailleBand' with haptic technology while preserving affordability. The `BrailleBand' enables passive reading using the Braille language. Connectivity between the BrailleBand and the smart device (phone) is established using Bluetooth protocol. It consists of six nodes in three bands worn on the arm to map the braille alphabet, which are actuated to give the sense of touch corresponding to the characters. Three mobile applications were developed for training the visually impaired and to integrate existing smart mobile applications such as navigation and short message service (SMS) with the device BrailleBand. %(eg:Facebook, twitter) with BrailleBand. 
The adaptability, usability and efficiency of reading was tested on a sample of blind users which reflected progressive results. Even though, the reading accuracy depends on the time duration between the characters (character gap) an average Character Transfer Rate of 0.4375 characters per second can be achieved with a character gap of 1000 ms. 
\end{abstract}

%%%%%%%%%%%%%%%%%%%%%%%%%%%%%%%%%%%%%%%%%%%%%%%%%%%%%%%%%%%%%%%%%%%%%%%%%%%%%%%%
\section{INTRODUCTION}

The global completely blind population is estimated to be 40 to 45 million and nearly 135 million are estimated to have low vision. %When compared to developed countries, developing countries account for 87 percent of the global blind population. 
The cause for blindness for youth is mainly due to birth defects in the brain or the eye, and uncorrected refractive errors. While visual impairment and blindness due to infections have significantly reduced with the rapid progress of health care services, there is a notable increase in blind and visually impaired people over 65 years of age due to long life expectancy. Unfortunately, the blind population is expected to double by 2020 \cite{world2014visual}. %\todo{Use a .bib file to maintain references. use \ ref{} to insert references - DONE}

Sight is the main human sense which possess the main influence on perception of all sensations, in collaboration with other senses such as hearing. Therefore, the lack of sight is the greatest challenge the blind face in performing their daily tasks such as navigation, information access, interpersonal interactions
%, communication, mobility 
and safety. Hence, the blind are unemployed and deprived of the privilege of education under normal circumstances. Approximately 75\%  of the blind are unemployed while only 10\% of the blind children receive special braille education \cite{worldMag}. In the context of the modern society a blind person and his/her family faces many socioeconomic problems. Consequently, the need for assistive technologies which enable the blind to live independent, productive and better lives emerged as investing on nursing homes, blind welfare, health care and blind care specialists were perceived to be costly and unsustainable solutions \cite{velazquez2010wearable}.
%\todo[inline]{Above para is too long. Reduce - DONE}

\subsection{Existing technologies and their drawbacks}

Many research projects have been conducted focusing on independent communication for the blind, the deaf and the deaf-blind communities. Wearable systems such as mechanical hands for automated fingerspelling and communication glove systems have been developed using various alphabets \cite{elsendoorn2004assistive}. The Lorm glove uses the Lorm alphabet, a form of communication for deaf and blind. It translates the hand-touch Lorm alphabet to text and vice versa using flat fabric sensors embedded on the glove \cite{gollner2012mobile}. DB-HAND is a similar wearable device with Malossi alphabet interface for the deaf-blind community. This glove is equipped with sensors and actuators enabling two way communication with pressing of tactile sensors enabling the inputs to the device \cite{caporusso2008wearable}. In another braille based mobile communication and translation glove, uses capacitive touch sensors and actuators to enable two way communication by translating text to braille patterns and vice versa \cite{choudhary2015braille}. Wearabraille is a keyboard for the deaf and blind which uses finger mount accelerometers to identify the tapping action of fingers in a similar fashion to typing in a traditional keyboard \cite{Wearabraille}.

However, the main drawback of glove systems is that they limit the use of hands for other activities since the sense of touch for communication is given to the palm or the fingers and it gets interrupted if hands are being used for another activity. 

Text-to-speech is another mode of assisting the blind. However, it possesses usability issues such as privacy issues, disturbs hearing which is the main sense of blind people etc. Even if a pair of headsets is used it permanently disconnects the visually impaired person from the surrounding environment increasing the chance of meeting with accidents. Moreover, Braille terminals and displays carry an exorbitant cost, hence unaffordable. 

\subsection{Proposed solution}

The hypothesis of the proposed solution is to use the sense of touch of the arm to transfer information to a blind person using a wearable band having six vibration nodes corresponding to the standardized braille dot code. The system developed to test this hypothesis integrated with a smart device was named as the `BrailleBand'. 

Wearing a band on the arm is similar to wearing a wristwatch, hence does not hinder the use of the hand for other essential tasks. Moreover, wearable assistive solutions are preferred than the portable assitive solutions such as braille displays and mobile phones (text to speech) as they provide hands-free convenient interactions.  

Assistive technologies basically provide the ability to disabled people to work independently without having a need for someone else attending. The main focus areas of assistive technologies are \cite{brabyn2007aids}:
\begin{enumerate}
\item Information transmission

Issues with regard to information transmission are reading, character identification and interpreting two and three-dimensional graphics.The most successful and most popular blind reading tool is braille dot code which is the same dot code used in the BrailleBand making it readily usable for the blind community for information transmission. Tactile displays and braille displays enable character and graphic recognition by transferring information to tangible forms to be felt by the blind. 

\item Mobility assistance/Navigation

Mobility and navigation involves information of the immediate dynamic environment and obstacle avoidance. The basic functionality of all electronic travel aids (ETAs) is scanning the surrounding environment and transferring the gathered information through other senses to the user. BrailleBand can transfer directional data through sense of touch to the user and give feedback on the traveling directions. 

\item Computer and smart device accessibility

With the increased use of computers and smart devices the need of the visually impaired to access smart devices emerged. The solutions currently leveraged are text-to-speech and braille displays. BrailleBand can transfer information on simple computer screens to blind users using the vibrating braille dot patterns and subsequent to the development of the replying system two way communication will be established with computers and smart devices. 

\end{enumerate}

\begin{figure}[thpb]
	\centering
	\includegraphics[width=8.9cm]{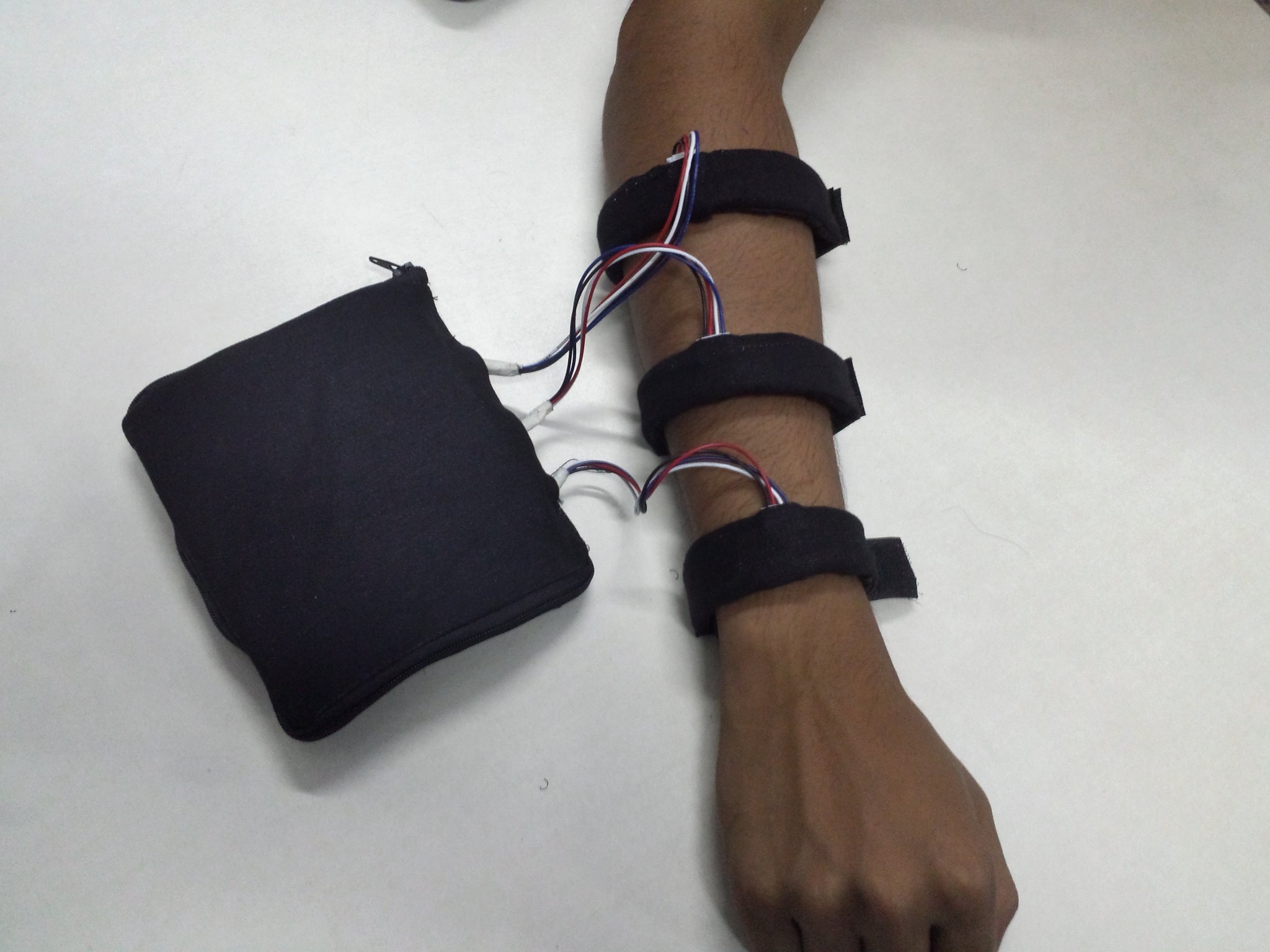}
    \caption{The BrailleBand}
      \label{fig:product}
\end{figure}

The product BrailleBand caters the needs of all the above three areas and its scope of applications are broader and can be customized according to diverse user needs (refer to Fig. \ref{fig:product}).

The study showed promising results with blind users being able to read mobile phone text messages and tweets using the BrailleBand. Also, the study revealed increase in reading speed and accurate character identification as the usage increased.% and with time they get more familiarized with the product which further enhances the . 

\section{METHODOLOGY}

\subsection{Hardware implementation}

\begin{figure} [thpb]
	\centering
    \includegraphics[width=8.9cm]{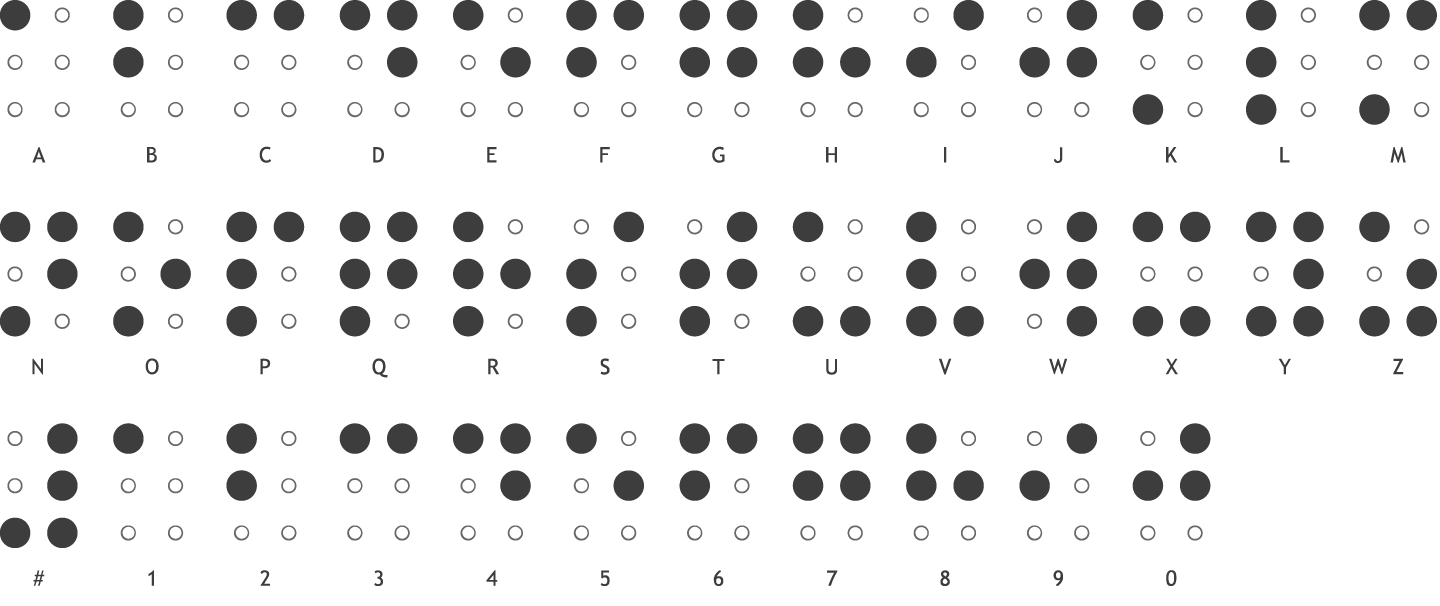}
	\caption{Braille alphabet and braille numbers (www.pharmabraille.com)} 
	\label{fig:dotcode}
\end{figure}

To represent nodes of the universal Braille dot code, six haptic motors (8mm Vibration Motor - 2mm Type from Precision Microdrives$^{TM}$) along with haptic motor controllers were used to generate the vibration sensation to be felt at the forearm corresponding to the characters to be read. Refer to Fig. \ref{fig:dotcode} for a few samples of Braille dot code.

\begin{figure} [thpb]
	\centering
    \includegraphics[width=8.9cm]{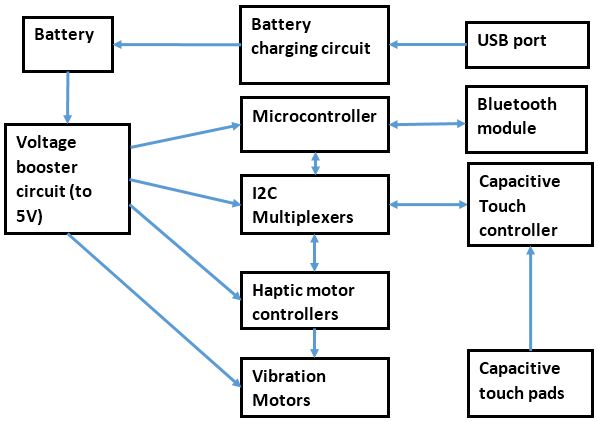}
    \caption{Interconnections and the functionality of the modules}
		\label{fig:HWIntFunc}
\end{figure}

The Fig. \ref{fig:HWIntFunc} summaries the hardware implementation of the Braileband. A Rechargeable 1200 mAh Li-Ion battery was used as the source of power to the whole system which through a 5V Voltage booster circuit. The battery was charged through the USB type-C port using a Battery Management 0.8A single input battery charger.

%fed drive microcontrollers, I2C (Inter-Integrated Circuit) Multiplexers, Motor controllers, capacitive touch controllers and vibration motors. 

The microcontroller used was ATMEGA328P-AU which interacts with the Bluetooth module and I2C multiplexers to generate the required signals corresponding to the information received from the smart device via Bluetooth.

Haptic motor controllers are I2C controllable. However, they have the same static I2C address and therefore I2C multiplexers were used.

%Haptic motor controllers and vibration motors (8mm Vibration Motor -2mm Type from Precision Mircodrives$^{TM}$) generate the vibration sensation to be felt on the hand of the user corresponding to the information which is to be delivered. BrailleBand has 6 vibrating nodes corresponding to the universal braille dot code and therefore when transmitting characters braille dot code is followed which familiarizes users to our product BrailleBand speedily. Refer Fig.4.

The circuit was designed using Altium$^{TM}$ and the enclosure of the prototype product was developed using Solidworks$^{TM}$ which was realized through three-dimensional (3D) printing. 

\subsection{Vibration motor placement and sense of touch}

When testing the hypothesis of this study, placement of vibration motors on the forearm was a critical task. Main aspects considered were:
\begin{enumerate}
\item Two-point discrimination threshold (TPDT)
\item Axes of vibration
\end{enumerate}

TPDT is a measure of the minimum distance between two pressure points on the skin to be felt as two distinct points. TPDT varies along body due to uneven distribution of mechanoreceptors. As BrailleBand is worn on the forearm the TPDT on the forearm which is 40mm is critical for design specifications \cite{amaralhandbook}.

The axis of vibration of the motors were horizontal and Eccentric Rotating Mass (ERM) haptic motors were used to generate vibrations. 

Even though the sense of touch is a supplementary sensory modality to a healthy-sighted person, it is the primary sensory modality for a blind person for non-audible information. Braille is a prime example of the success of tactile mode of transferring information. 

The skin is responsible for the sense of touch through three types of sensory modalities:
\begin{enumerate}
\item Thermoreceptors - for thermal sensations
\item Nociceptors - pain sensing
\item Mechanoreceptors - mechanical sensations due to external forces
\end{enumerate}

There are four types of mechanoreceptors in the human skin and out of them Meissner senses touch while Pacini corpuscles sense vibration. Since the BrailleBand stimulus is vibration our interest was on Meissner and Pacini sensors \cite{sekuler2002perception}.

\begin{itemize}
\item Meissner - high amplitude low frequency stimuli
\item Pacini corpuscles - low amplitude high frequency stimuli
\end{itemize} 

The frequencies and the amplitudes of the haptic motors were determined in such a manner that they stimulate both Pacini and Meissner vibration sensors in the skin to enhance the sense of vibration. 

\subsection{Software implementation}

\subsubsection{Connectivity}

Mobile applications which are connected with the device are developed using android. By selecting the Media Access Control (MAC) address of the Bluetooth module of the device (HC06), application can establish a socket connection between the BrailleBand and the smart phone. After the establishing the connection, a thread is created in the application which handles the socket connection. When the thread is started, the input stream and the output stream is set with the BrailleBand. The gaps between characters and words are controlled by changing the threads to postDelayed state. %Here we have programmed ATMEGA 328P microcontroller by the Arduino software. 
The program continuously read the input stream and control haptic motors depending on the received character. 

\subsubsection{Mobile applications}

Three mobile applications were developed for the BrailleBand:
\begin{enumerate}
\item BrailleBand Teacher

Used to get the blind person familiarized with the device and to teach the method of reading braille from BrailleBand. %Since, blind people are deprived of their sight, the other senses are stronger and hence they catch up the system really fast. 

\item BrailleBand Messenger

Enables the user to read text messages received to the mobile phone through the BrailleBand. 

\item BrailleBand Navigator

This application can navigate a blind person to a destination by giving haptic feedback about the direction and the distance to travel through the BrailleBand. However, the accuracy of navigation still needs to be developed. 

%\todo[inline]{write the application to the blind - DONE}
%Navigator application is used to navigate not only blind people but also to navigate bike riders as they cannot use the mobile phone while riding bikes. 

\end{enumerate}

\begin{figure}[thpb] 
	\centering
    \includegraphics[width=8.9cm]{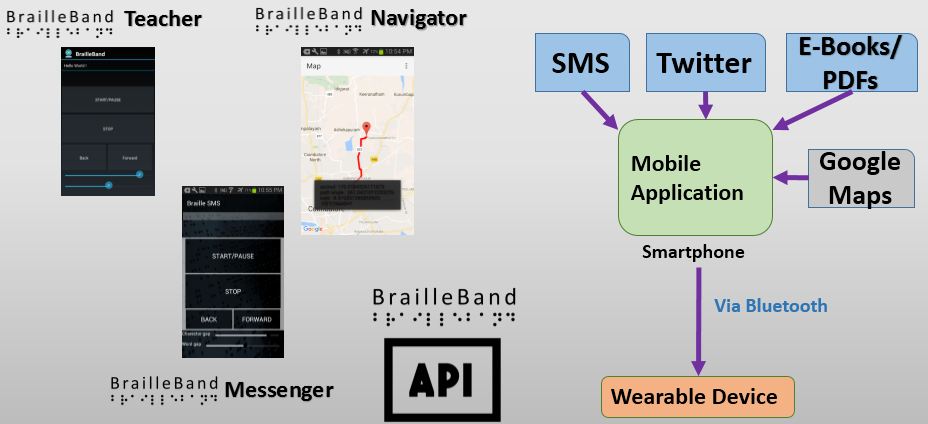}
    \caption{BrailleBand mobile applications and integration of existing applications with product BrailleBand}
    \label{figure_apps} 
\end{figure}

The existing social media and utility mobile applications such as facebook, twitter, linkedin and google maps can be integrated with the BrailleBand application and is readily usable with the BrailleBand device since information from these applications can be conveyed to the blind user through the device. Refer to Fig.\ref{figure_apps}.

An android library was also developed and is readily available at \textit {https://github.com/BrailleBand/BrailleBand-mobile-app} to facilitate other developers to develop mobile applications customized to different user requirements. Hence, the product BrailleBand is open for innovations. 

\subsubsection{Microcontroller programming}
The microcontroller is programmed using Arduino software to convert the character which is being received from the bluetooth module to the respective braille pattern. Then using the I2C communication protocol the microcontroller addresses the desired haptic motors and activate them. 

\subsection{Testing and verification}

Three experiments were conducted to test and validate the hypothesis which resulted in the following parameters:
\begin{enumerate}
\item Reading speed 
\item Character transfer rate (CTR)
\item Device usability score
\end{enumerate} 

A sample of 10 blind subjects (7 male and 3 female) were chosen from the Blind School, Rathmalana, Sri Lanka within an age range of 16 to 19 years based on the criteria of having experience with Braille language for more than 5 years. No initial training was given to the students but 15 minutes were taken to familiarize the students with the device using the BrailleBand Teacher mobile application. The sample was tested with the device where 3 character words were sent with different character gaps. The critical parameter which governs the Reading Speed and the CTR is the Character gap and hence we focused on reducing the character gap and identifying the minimum character gap which does not significantly differ statistically with the character gap which gives the best average reading accuracy (90.1\%) using Analysis of Variance (ANOVA), Bonferroni and Holm tests. It was found that the minimum character gap satisfying the above conditions is 1000 ms. Refer to Table \ref{table_1}.  

\begin{equation}
CTR = \frac{number\, of\, characters\, sent}{time\, taken}
\label{eqn_1}
\end{equation}

An exact CTR cannot be calculated since the number of dots to be vibrated vary with characters. Hence, a minimum (for ``q") and a maximum (for ``a") CTR was calculated and a average CTR was derived by using (\ref{equ:CTRavg}). 

\begin{equation}
CTR_{average} = \frac{{CTR_{maximum}} + {CTR_{minimum}}}{2}
\label{equ:CTRavg}
\end{equation}

The subject was given a score ranging from 0 to 10 to rate the usability of the BrailleBand based on the preference to adopt and use this device for communication purposes.

\section{RESULTS}

\subsection{Development of the BrailleBand}
BrailleBand is a blind support wearable which is light weight, portable and easy to use. Moreover, it mitigates the usability and affordability issues associated with the current solutions available. Blind support wearable solutions are still progressing in the experimental stages and hence the market is open with minimum competition. BrailleBand caters a broader scope of user needs such as information transmission, navigation and mobility assistance, accessibility to computers and smart devices.

\subsection{Reading speed test} 
The sample was tested with random words consisting of three characters and the character gap was reduced as in Table I and the following reading accuracies were noted. With decreasing character gap the reading speed increases as characters are being transmitted faster. 

\begin{table}[h]
\caption{Reading speed test results}
\label{table_1}
\begin{center}
\begin{tabular}{|c|c|c|}
\hline
Character gap(ms) & Average reading accuracy(\%) & Standard deviation \\
\hline
2000 & 90.1 & 10.94\\
\hline
1500 & 90.1 & 6.24\\
\hline
1200 & 87.9 & 6.24\\
\hline
1000 & 87.9 & 12.11\\
\hline
800 & 69.1 & 18.03\\
\hline
500 & 53.2 & 21.02\\
\hline
400 & 46.4 & 22.84\\
\hline
\end{tabular}
\end{center}
\end{table}

It is notable that with practice the reading speed can be improved to a greater extent. 

\subsection{Character Transfer Rate (CTR)}
A dot in the braille cell is vibrated for a period of 300 ms and the gap between the vibration of two consecutive dots is also 300 ms. Therefore, for a single dot to be vibrated it will take 600 ms. 

An ANOVA test can be done to check whether the average reading accuracies obtained in Table I are statistically significantly different with a confidence level of 95\%.

\begin{table}[h]
\caption{One-way ANOVA for 7 independent character gaps}
\label{table_2}
\begin{center}
\begin{tabular}{|c|c|c|c|c|c|}
\hline
source & sum of squares & df & mean square & F statistic & p-value\\
\hline
treatment & 21,168.37 & 6 & 3,528.06 & 15.1239 & 0\\
\hline
error & 14,696.5 & 63 & 233.28 & - & - \\
\hline
total & 35,864.87 & 69 & - & - & - \\
\hline
\end{tabular}
\end{center}
\end{table}

The conclusion of the ANOVA test as tabulated in Table II is that the p-value corresponding to the F-statistic of one-way ANOVA is lower than 0.05 (95\% confidence), suggesting that the sample data of one or more treatments (character gaps) are significantly different to the others. Therefore, a multiple comparison method should be used to assess the statistical significance of the differences between the means of the sample data under different treatments. 

Hence, Bonferroni and Holm multiple comparison tests follow to test only pairs relative to a certain treatment (character gap) to verify statistical significant differences. 
Since the character gap 1500 ms gives the “highest reading accuracy (90.1\%)” while having the least standard deviation (refer to Table I), we choose character gap = 1500 ms as the treatment against which other treatments are compared for statistical significant differences. 

\begin{table}[h]
\caption{Bonferroni and Holm results: only pairs relative to Character gap=1500 {ms} simultaneously compared}
\label{table_3}
\begin{center}
\begin{tabular}{| p{2cm} | p{2cm} | p{1.5cm}|p{1.5cm}|}
\hline
Treatment pairs & Bonferroni and Holm T-statistic & Bonferroni inference & Holm inference\\
\hline
1500 Vs 2000 & 0.00 & Insignificant & Insignificant \\
\hline
1500 Vs 1200 & 0.32 & Insignificant & Insignificant \\
\hline
1500 Vs 1000 & 0.32 & Insignificant & Insignificant \\
\hline
1500 Vs 800 & 3.07 & Insignificant & Significant \\
\hline
1500 Vs 500 & 5.40 & Significant & Significant \\
\hline
1500 Vs 400 & 6.40 & Significant & Significant \\
\hline
\end{tabular}
\end{center}
\end{table}

From the above results table of Bonferroni and Holm multiple comparison tests, we can observe that there is statistically insignificant difference between the sample data under Character Gap=1500 ms, Character Gap=2000 ms, Character Gap=1200 ms, and Character Gap=1000 ms. 

Therefore, we can use the character gap of 1000 ms for Character Transfer Rate (CTR) calculations as the sample data do not show a statistically significant difference with the treatment which gives the highest reading accuracy of 90.1\% (Character Gap=1500 ms). 

Therefore, we can calculate the maximum CTR which is corresponding to the character "a" and minimum CTR corresponding to the character "q" using Equation \ref{eqn_1}. 

$$CTR_{maximum} = \frac{1}{0.6 + 1} = 0.625\,characters\,s^{-1}$$ 

$$CTR_{minimum} = \frac{1}{0.6\times 5 + 1} = 0.25\,characters\,s^{-1}$$

Therefore, from Equation 2, 
$$CTR_{average} = \frac{0.625 + 0.25}{2} = 0.4375\,characters\,s^{-1}$$

% \begin{table}[h]
% \caption{Reading accuracy test results}
% \label{table_example}
% \begin{center}
% \begin{tabular}{|c|c|}
% \hline
% Hours of Training & Average number of misreads\\
% \hline
% 0 & 5.0\\
% \hline
% 3 & 4.2\\
% \hline
% 6 & 3.6\\
% \hline
% 9 & 3.1\\
% \hline
% 12 & 2.6\\
% \hline
% 15 & 2.1\\
% \hline
% \end{tabular}
% %\end{center}
% %\end{table}

It is observed that the average CTR is 0.4375 characters per second and this also can be improved with practise. 

\subsection{Usability score} 
All the subjects were requested to rate the usability of the BrailleBand out of 10 and the resultant average usability score was 8.7.

\section{DISCUSSION}

BrailleBand caters a wider range of blind support needs and it has many applications such as reading, navigation and smart device accessibility. It is a complete package of the device and the mobile applications for smart phones. 

However, the replying system of the BrailleBand is still under development. The capacitive touch pads and capacitive touch controllers are the constituents of the replying system which allows the user to reply back to messages using the same braille dot pattern. Refer to Fig. \ref{fig:HWIntFunc}. 

Additionally, further research should be done on the health and safety aspect of the product as well. The go-to-market plan should include a market research to identify the additional features demanded by the users and to identify the price the users are willing to pay. Moreover, long term sustainable usability of the product should also be verified as BrailleBand will be a lifetime product for a blind person focusing on robustness of the device and biological effects as well (for example: nerve adaptations for vibrations).

The reading speed and the CTR can be further improved with practise. Moreover, from the device end we can reduce the dot vibration time which is currently at 300 ms to a lesser value and transmit characters faster. But the compromise here is the effect of the vibrating sensation to the user. 

As blind people usually lack financial strength to afford sophisticated devices. Therefore, the BrailleBand will be a better solution. Yet, the price of BrailleBand should be subsidized  through corporate and government funding to reach the wider blind community.

Wearable assistive technology solutions for blind are still at an emerging level with lot of research and development being done around the world. BrailleBand integrates haptics in developing an assistive device for blind support.  

\section{CONCLUSIONS}

The blind community uses the sense of hearing and sense of touch to interact and understand the surrounding environment. Hence, the sense of touch becomes the primary sensory modality to communicate non-audible information to and from a blind person.

The `BrailleBand' haptic wearable blind support device connected to smart phone applications helps the blind community to lead an independent quality life. This assistive technology enables information transmission, navigation and smart device accessibility through the sense of touch. With the development of  the BrailleBand we have successfully implemented a new mode of communication to the blind using haptic technology. 

The product feasibility tests conducted showed promissing results with regard to reading speed and reading accuracy which improved with practice. Based on experimental results an average CTR of 0.4375 characters per second was achieved. 

 %Based on this assumption the BrailleBand uses haptic technology, which is the science of applying touch sensation to communicate information. 

\addtolength{\textheight}{-12cm}   % This command serves to balance the column lengths

%%%%%%%%%%%%%%%%%%%%%%%%%%%%%%%%%%%%%%%%%%%%%%%%%%%%%%%%%%%%%%%%%%%%%%%%%%%%%%%%

%%%%%%%%%%%%%%%%%%%%%%%%%%%%%%%%%%%%%%%%%%%%%%%%%%%%%%%%%%%%%%%%%%%%%%%%%%%%%%%%

%%%%%%%%%%%%%%%%%%%%%%%%%%%%%%%%%%%%%%%%%%%%%%%%%%%%%%%%%%%%%%%%%%%%%%%%%%%%%%%%
\section*{ACKNOWLEDGMENT}
We extend our gratitude to Dr. Ajith Pasqual and all the academic staff of Department of Electronic and Telecommunication Engineering, University of Moratuwa for veteran assistance and guidance given in developing the product BrailleBand. 
We acknowledge the Blind School, Rathmalana, Sri Lanka for accommodating the testing and verification process by granting patient access. 

%%%%%%%%%%%%%%%%%%%%%%%%%%%%%%%%%%%%%%%%%%%%%%%%%%%%%%%%%%%%%%%%%%%%%%%%%%%%%%%%
\bibliographystyle{plain}
\bibliography{ref}
\nocite{*}

\end{document}